\documentclass[aps,showpacs,amssymb,%
floatfix,superscriptaddress]{revtex4}
\usepackage{graphicx}
\usepackage{epsf}
\usepackage{dcolumn}
\usepackage{bm}
\usepackage{hyperref} 
\usepackage{latexsym}
\usepackage{amsmath}

\begin{document}
\def\av#1{\langle#1\rangle}
\def\etal{{\it et al\/.}}
\def\ie{{\it i.e.}}
\def\prob{{\rm Prob}}
\def\remark#1{{\bf *** #1 ***}}

\title{On the Relation between One-Species Diffusion-Limited Coalescence and Annihilation in One Dimension} 
\author{Daniel~ben-Avraham} 
\affiliation{Department of Physics, Clarkson University,
Potsdam NY 13699-5820, USA}
\author{\'Eric~Brunet}
\affiliation{\'Ecole Normale Sup\'erieure, 24 rue Lhomond,
75230 Paris Cedex 05, France}

\begin{abstract}
The close similarity between the hierarchies of multiple-point
correlation functions for the diffusion-limited coalescence and
annihilation processes has caused some recent confusion, raising doubts
as to whether such hierarchies uniquely determine an infinite particle
system.  We elucidate the precise relations between the two processes,
arriving at the conclusion that the hierarchy of correlation functions
does provide a complete representation of a particle system on the line.
We also introduce a new hierarchy of probability density functions, for
finding particles at specified locations and none in between.  This
hierarchy is computable for coalescence, through the method of empty
intervals, and is naturally suited for questions concerning the ordering of 
particles on the line.
\end{abstract}
\maketitle

Diffusion-limited coalescence, $A+A\to A$, and annihilation, $A+A\to0$,
on the line, have long been known to display anomalous kinetics
(different from the mean-field reaction-limited regime) and to belong to
the same universality class~\cite{Bramson80}-\cite{Derrida96}. In fact,
the similarities run deeper than that, as the full hierarchy of
multiple-point correlation functions in the two processes, when expressed
in  different length scales, are
identical~\cite{Henkel95,Krebs95,Balboni95,Simon95,Masser01}.  Yet there
exist important differences between the two processes.  Most
conspicuously, the density function for the gap $x$ between adjacent
particles (the so called inter-particle distribution function, or IPDF)
falls off, for $x\to\infty$, as $e^{-\alpha x^2}$ for coalescence, but only
as $e^{-\beta x}$ for annihilation~\cite{Bramson80,dba90,Alemany95,Derrida96}.
This
raises the question whether the infinite hierarchy of multiple-point
correlation functions uniquely determines a system of particles on the
line~\cite{Masser01}.

In this communication, we answer this question on the affirmative: the
hierarchy of multiple-point correlation functions does provide a unique
representation of an infinite particle system.  Indeed, the IPDF can be
computed from the multiple-point correlation functions using an
inclusion-exclusion formula, as shown in Eq.~(\ref{method1}). The correlation
functions of the two processes are simply not the same, despite the
similarity upon the rescaling of space. The precise relationship
between coalescence and annihilation, and its consequences, is thoroughly
discussed. An alternative hierarchy of probability density functions,
for finding particles at locations $x_1,\dots,x_n$ {\em and none in
between}, is introduced, and we show how to compute it explicitly, for
coalescence, using the method of empty
intervals~\cite{dba90,doering91,doering92,dba95,dba98,Khorrami01}. 
This new hierarchy is naturally suited for answering questions concerning
the ordering of particles on the line.  It is further used
for obtaining an exact new expression, as well a series of systematic
approximations for the IPDF of the annihilation process~\cite{Alemany95,Derrida96}.

\section*{Equivalence between coalescence and annihilation}

The $n$-point correlation functions (the joint
probability for finding particles at positions~$x_1,
\dots, x_n$ at time $t$, simultaneously) of diffusion-limited annihilation and
diffusion-limited coalescence in one dimension seem very similar.  Indeed, 
for suitable initial conditions, one  has
the exact result~\cite{Henkel95,Krebs95,Balboni95,Simon95,Masser01}:
\begin{equation}
\rho_n^{\text{anni}}(x_1,\dots,x_n;t)
	=\frac{1}{2^n}\rho_n^{\text{coal}}(x_1,\dots,x_n;t).
\label{equiv}
\end{equation}
The relation~(\ref{equiv}) has the following simple interpretation:  {\em To obtain a
configuration of the annihilation process, with the correct weight, select
a configuration of the coalescence process but then retain only {\em half} of the
particles, {\em i.e.}, randomly, and independently, retain (or discard) each of the original 
particles with probability 1/2}.

This interpretation follows from the well-known observation that the two
processes may be realized simultaneously.  Starting from a random
initial configuration of particles on the line, tag each of the particles
with  probability~$1/2$ and run a diffusion-limited coalescence process: 
When two particles meet they coalesce into a single untagged
particle if the parents are alike (either normal or
tagged), and into a tagged particle if the parents are different.
Clearly, the set of all particles in the system (tagged and untagged) represents
a configuration of the coalescence process $A+A\to A$, while the subset of
tagged particles  follows the annihilation process $A+A\to0$ . 
At any time~$t$ a particle is untagged if and only if it has an even number of tagged 
ancestors at time~$t=0$, while the ancestors of two different particles form disjoint
sets.  It follows that at time $t$ particles are tagged with probability $1/2$, independently 
from one another.

\medskip

For the suitable initial conditions for which (\ref{equiv}) holds, the
density of particles in the anihilation process is half the density in
the coalescence process. It is tempting to rescale space in the
coalescence process by a factor of two, to impose the same density in both
processes. Defining~$y_i=2x_i$ and effecting the change of variables $\rho_n^{\text{coal}}(x_1,\dots
x_n;t)\to\tilde\rho_n^{\text{coal}}(y_1,\dots
y_n;t)$, one gets
\begin{equation}
\rho_n^{\text{anni}}(x_1,\dots,x_n;t)
	=\tilde\rho_n^{\text{coal}}(y_1,\dots,y_n;t).
\label{equiv2}
\end{equation}
This relation has caused some confusion, leading one of us (DbA) to erroneously conclude that since
the distribution of particles in coalescence and annihilation are different (as evidenced,
for example, from their different inter-particle gap distribution functions) it must be the case
that the infinite hierarchy of multiple point correlation functions does not uniquely determine
an infinite set of points on the line~\cite{Masser01}.
Eq.~(\ref{equiv2}) does not mean, however, that the functions
$\rho_n^{\text{anni}}$ and $\tilde\rho_n^{\text{coal}}$ are identical: they
are applied to different arguments.   Indeed, the stretching of space by a factor of two
and eliminating half of the particles at random are {\em not} generically equivalent~\cite{poisson}.
An easy way to see this is by considering the effect of the two operations on a lattice of equally
spaced particles: stretching yields a new lattice with double the gap between particles, whereas
the random elimination of half the particles leads to a disordered array.

\section*{Inter-particle distribution functions (IPDF)}

The hierarchy of multiple-point correlation functions determines the system uniquely.
Some quantities, however, such as the IPDF, are notoriously difficult to obtain, using the $\rho_n$.
This quantity is
readily available for the coalescence process, through the method of
empty intervals, but not for the annihilation process.

To obtain the IPDF, one needs to compute the probability density
$P_0(x_1,x_2)$ for finding particles at $x_1$ and $x_2$ and \emph{no
particles in between}. The density probability function of the gap~$x$
between particles is then $P_0(0,x)/\rho$~\cite{time}.
$P_0(x_1,x_2)$ is, in principle, available from the $\rho_n$. Indeed,
\begin{equation}
P_0(x_1,x_2)=\rho_2(x_1,x_2)-\int_{x_1}^{x_2} dz_1\ \rho_3(x_1,z_1,x_2)
+ \int_{x_1}^{x_2}dz_1\int_{z_1}^{x_2}dz_2\ \rho_4(x_1,z_1,z_2,x_2)-\dots.
\label{method1}
\end{equation}
On the right hand side of this equation, 
events with exactly one particle
between~$x_1$ and $x_2$ are counted once by the first term but cancelled
by the second term; events with two particles in between are counted once
each by the first and third terms, but are cancelled by the second term,
which counts these events twice ($z_1$ representing the first or second
intervening particle), etc.  In this fashion, only the events with no
particles between~$x_1$ and $x_2$ are accounted for at the end. 

More generally, one can write down a similar expression for
$P_k(x_1,x_2)$ --- the probability density for finding particles at $x_1$
and $x_2$ and exactly $k$ particles in between:
\begin{equation}
P_k(x_1,x_2)=\sum_{n\ge k}(-1)^{n-k}\binom{n}{k}R_n(x_1,x_2),
\label{PfromRho}
\end{equation}
where
\begin{equation}
R_n(x_1,x_2)=\!\!\!\!\!\idotsint\limits_{x_1<z_1<\dots<z_n<x_2}\rho_{2+n}(x_1,z_1,\dots,z_n,x_2)
\;dz_1\cdots dz_n.
\end{equation}

In principle, all the~$\rho_n$ can be computed explicitly, both in the
annihilation and the coalescence process, but their actual expressions are
complicated and (\ref{method1}) becomes impractical.   Another
possibility is to derive $P_0^\text{anni}$ from the~$P_k^\text{coal}$
using the correspondence between the configurations of both processes:  To
obtain a configuration of the anihilation process with particles at
positions~$x_1$ and $x_2$ and nothing in between, we start with a
configuration of the coalescence process with particles at~$x_1$ and
$x_2$ and exactly $k$ other particles in between.  Retaining each of these
particles with probability $1/2$, there is a probability~$1/4$ that
the particles at~$x_1$ and $x_2$ stay put, and a probability~$(1/2)^k$ to kill
the~$k$ particles in between. Therefore, we have
\begin{equation}
P_0^\text{anni}(x_1,x_2)={1\over4}\sum_{k\ge0}{1\over2^k}P_k^\text{coal}(x_1,x_2).
\label{method2}
\end{equation}
This is but a special case of Eq.~(23) in~\cite{Derrida96}, already suggested by Derrida and Zeitak.

Obtaining the $P_k^\text{coal}(x_1,x_2)$ from the $\rho_n^{\text{coal}}$,
remains still an impractical proposition.
Fortunately, in the case of the coalescence
process, the method of intervals allows  for a more direct way.

\section*{New hierarchy for finding $n$ sequential particles}

Dimension 1 is special in that one can meaningfully discuss
the ordering of the particles 
on the line.  Thus, instead of the traditional hierarchy of multiple-point
correlation functions, we propose an hierarchy designed to keep track of the sequential order
of the particles.

Let $\omega_n(x_1,\dots,x_n;t)$ denote the probability for finding particles at $x_1,x_2,\dots,x_n$
at time $t$, but no other particles
in the intervals $(x_1,x_2),(x_2,x_3),\dots,(x_{n-1},x_n)$.
Clearly, the $\omega_n$ determine a distribution uniquely.  Indeed, 
\begin{eqnarray}
\rho_2(x_1,x_2)&=&\sum_{n=0}^{\infty}\>\;\idotsint\limits_{x_1<z_1<\dots<z_n<x_2}\omega_{2+n}(x_1,z_1,\dots,z_n,x_2)
\;dz_1\cdots dz_n,\\
\rho_3(x_1,x_2,x_3)&=&\sum_{k,l}\>\;\idotsint\limits_{x_1<y_1<\dots<y_k<x_2<z_1<\dots<z_l<x_3}
\!\!\!\!\!\!\!\!\omega_{3+k+l}(x_1,y_1,\dots,y_k,x_2,z_1,\dots,z_l,x_3)
\;dy_1\cdots dy_k\,dz_1\cdots dz_l,
\end{eqnarray}
and similar expressions for $\rho_m$, $m>3$.  Thus, defining $\omega_1(x)\equiv\rho_1(x)$,
the complete hierarchy of multiple-point correlation functions, $\{\rho_n\}$, can be derived from the hierarchy of sequential particles, $\{\omega_n\}$.

The $\omega_n$ are better suited to deal with questions regarding ordered sets of particles.  A relevant example are the $P_k$, which instead of the infinite sum in~(\ref{PfromRho}) are now simply given by
\begin{equation}
P_k(x_1,x_2)=\!\!\!\!\!\idotsint\limits_{x_1<z_1<\dots<z_k<x_2}\omega_{2+k}(x_1,z_1,\dots,z_k,x_2)
\;dz_1\cdots dz_k,
\label{PfromOmega}
\end{equation}
and, in particular, $P_0(x_1,x_2)=\omega_2(x_1,x_2)$.

In the case of coalescence,  the hierarchies $\{\rho_n\}$ and
$\{\omega_n\}$
 may be derived explicitly through the method of empty intervals.  Specifically, the method of intervals yields expressions for
$E_n(x_1,y_1,\dots,x_n,y_n;t)$ --- the probability that the intervals $(x_1,y_1),\dots,(x_n,y_n)$ be simultaneously empty at time $t$~\cite{dba98}.  The two hierarchies are obtained as spatial derivatives of the $E_n$, but evaluated at different points:
\begin{eqnarray}
\label{omega1}
\omega_n(x_1,\dots,x_n;t)&=&
\frac{\partial^n}{\partial x_1\cdots\partial x_n} E_{n}(x_1,y_1,\dots,x_n,y_n;t)|_{y_1=x_2,
  \dots,y_{n-1}=x_n,y_n=x_n},\\
\label{rho1}
\rho_n(x_1,\dots,x_n;t)&=&
\frac{\partial^n}{\partial x_1\cdots\partial x_n} E_n(x_1,y_1,\dots,x_n,y_n;t)|_{y_1=x_1,
  \dots,y_n=x_n}.
\end{eqnarray}
Actually, $\omega_n$ can be computed somewhat more cheaply, from $E_{n-1}$ rather than $E_n$:
\begin{equation}
\omega_n(x_1,\dots,x_n;t)=
-\frac{\partial^n}{\partial x_1\cdots\partial x_{n-1}\partial y_{n-1}} E_{n-1}(x_1,y_1,\dots,x_{n-1},y_{n-1};t)|_{y_1=x_2,
  \dots,y_{n-1}=x_n}.
\label{omega2}
\end{equation}
We have written Eq.~(\ref{omega1}) merely to showcase the beautiful symmetry between $\rho_n$ and $\omega_n$. 

The first few $\omega_n^{\text{coal}}$ computed for coalescence, in the long time asymptotic limit, using the empty intervals derived in~\cite{dba98,Masser01}, are~\cite{Derrida96}
\begin{eqnarray}
\omega_2^{\text{coal}}(x_1,x_2;t)=&&\sqrt{\pi}\rho^2\xi_{12}e^{-\xi_{12}^2},\\
\omega_3^{\text{coal}}(x_1,x_2,x_3;t)=&&\sqrt{\pi}\rho^3\xi_{13}(e^{-\xi_{12}^2-\xi_{23}^2}-e^{-\xi_{13}^2}),\\
\omega_4^{\text{coal}}(x_1,x_2,x_3,x_4;t)=&&\sqrt{\pi}\rho^4\{
    \xi_{14}(e^{-\xi_{14}^2}-e^{-\xi_{14}^2-2\xi_{23}^2}+e^{-\xi_{12}^2-\xi_{23}^2-\xi_{34}^2}
    -e^{-\xi_{12}^2-\xi_{24}^2}+e^{-\xi_{13}^2-\xi_{23}^2-\xi_{24}^2}-e^{-\xi_{13}^2-\xi_{34}^2})\nonumber\\
    &&+\sqrt{\pi}(\xi_{14}\xi_{23}e^{-\xi_{14}^2-\xi_{23}^2}
    -\xi_{13}\xi_{24}e^{-\xi_{13}^2-\xi_{24}^2}+\xi_{12}\xi_{34}e^{-\xi_{12}^2-\xi_{34}^2})
    \text{erfc}(\xi_{23})\},
\end{eqnarray}
where $\rho=1/\sqrt{2\pi Dt}$ is the long time asymptotic density of particles in the coalescence process,
$\xi_{ij}\equiv(x_j-x_i)/\sqrt{8Dt}$, and $\mathrm{erfc}(x) = (2/\sqrt\pi) \int_x^{\infty} \exp(-u^2)\,du$ is
the complementary error function.  


\medskip

Returning to the question of the IPDF in the annihilation process,
all the $\omega_k^\text{coal}$ can be computed, in principle, but
obtaining the exact $P_0^\text{anni}$ from (\ref{PfromOmega}) and
(\ref{method2}) is not an easy task. Nevertheless, this approach can be
used to generate efficient approximations of the IPDF.
For instance, for small inter-particle gaps~$x=x_2-x_1$, 
one need only keep
the first few terms in (\ref{method2}), since the probability of finding several particles in the 
gap becomes negligibly smaller as their numbers increase.
Indeed, 
using just the first term in (\ref{method2}) yields an expression
that matches the exact result (Eq~(43) in \cite{Derrida96}) to order $x^3$;
the first two terms improve the match up to order $x^7$, and three terms up to $x^{12}$.
However, truncating~(\ref{method2}) in this fashion, at any order, gives terrible results for large $x$, as it 
predicts a Gaussian decay, $\exp(-\pi x^2)$, instead of the correct
exponential decay.

For large values of $x$, the simplest assumption that the gaps between
particles are independent,
\begin{equation}
\omega_{2+k}^{\text{coal}}(x_1,z_1,z_2,\dots,z_k,x_2)\approx
\frac{\omega_2^{\text{coal}}(x_1,z_1)\omega_2^{\text{coal}}(z_1,z_2)\cdots \omega_2^{\text{coal}}(z_k,x_2)}%
{\rho^k}.
\label{approx1}
\end{equation}
leads to to an exponential decay of the IPDF, $P_0^\text{anni}(0,x)\simeq 1.6777 \exp(-1.2685 x)$, at
large~$x$ (where distance is scaled so that the density of particles is equal to 1).
The same result was already derived
in~\cite{Alemany95}, using a similar assumption of uncorrelated
particles, and compares favorably with the exact result of
$P_0^\text{anni}(0,x)\sim1.8167\exp(-1.3062 x)$ \cite{Derrida96}.
Systematic improvements are achieved by taking into account more correlations. 
For instance, assuming
\begin{equation}
\omega_{2+l}^{\text{coal}}(x_1,z_1,z_2,\dots,z_{l},x_2)\approx
\frac{\omega_3^{\text{coal}}(x_1,z_1,z_2)\omega_3^{\text{coal}}(z_2,z_3,z_4)\cdots
\omega_3^{\text{coal}}(z_{l-1},z_{l},x_2)}{\rho^l}
\label{approx2}
\end{equation}
(for $l$ odd; for $l$ even, the last term in the product is
$\omega_2^\text{coal}(z_l,x_2)$),
leads to $P_0^\text{anni}(0,x)\simeq
1.728976 \exp(-1.285339 x)$.


\section*{Discussion}

In summary, we have elucidated the exact relation between
diffusion-limited one-species coalescence and annihilation in one
dimension.  The precise meaning of the similarity between the respective
hierarchies of multiple-point correlation functions, Eq.~(\ref{equiv}),
has been clarified: configurations of the annihilation process are
obtained by random elimination of half the particles in the coalescence
process.  This, however, is {\em not} equivalent to the stretching of
space by a factor of two.  

We have also introduced a different hierarchy of probability density functions: the
$\omega_n(x_1,\dots,x_n;t)$ --- the probability for finding
particles at  $x_1,\dots,x_n$ and no particles in between, at time $t$.
This new hierarchy capitalizes on the topological constraints special to one
dimension, and is better suited for answering questions concerning \emph{specific} numbers
of particles.

Both the traditional multiple-point correlation functions, the $\rho_n$, and the $\omega_n$
can be derived from the distribution of empty intervals on the line, in a way that highlights 
their relationship, Eqs.~(\ref{omega1}), (\ref{rho1}).  The two hierarchies determine an
infinite system of particles on the line completely, and in particular it is possible to express
the $\rho_n$ using the $\omega_n$, and vice-versa, albeit at the cost of resorting to infinite
sums of unwieldy integrals.  In this sense, the best situation occurs when the distribution
of empty intervals is available, for it yields the $\rho_n$ and $\omega_n$ directly.
This is the case for the coalescence process, but not for annihilation.  
For annihilation the $\rho_n$ are known, but the $\omega_n$ are harder to
obtain.
Derrida and Zeitak have obtained the IPDF (directly related to $\omega_2$) exactly~\cite{Derrida96}.
Here we presented an alternative approach, based on the relation between the $\omega_n$ and $\rho_n$, that provides useful approximations.

The situation is not yet completely clear even for coalescence. For example, consider the probability
of finding two particles separated by a distance $x$ and having exactly $k$ particles in between,
$P_k^{\text{coal}}(0,x)$.  For small gaps, $x\ll1/\rho$, we obtain $P_k^{\text{coal}}(0,x)\sim x^{\alpha(k)}$,
with $\alpha(k)=1$, $4$, $8$, $13$, for $k=0$, $1$, $2$, $3$, respectively.  We have not yet found a satisfactory way to predict $\alpha(k)$ for arbitrary $k$.  It remains the subject of future studies.


\acknowledgments
We are grateful to the organizers of the Dresden 2003 workshop on
\emph{Non-equilibrium Statistical Physics in Low Dimensions and
Reaction-Diffusion Systems} for giving us the opportunity to meet and
discuss,
and to NSF grant PHY-0140094 (DbA) for partial support of this research.


\begin{thebibliography}{99}
\bibitem{poisson} A notable exception is when the particles are distributed randomly, in a Poisson distribution.
\bibitem{time} Here, and elsewhere in the text, the missing time argument $t$ is implied.
\end{thebibliography}
\end{document}